# Improving SABRE hyperpolarization with highly non-intuitive pulse sequences: moving beyond avoided crossings to describe dynamics

Shannon L. Eriksson,[1,2] Jacob R. Lindale,[1] Xiaoqing Li[3], Warren S. Warren,[1,2,3,4*]

[1]Department of Chemistry, Duke University, Durham, NC, 27708
[2]School of Medicine, Duke University, Durham, NC, 27708
[3]Department of Physics, Duke University, Durham, NC, 27708
[4]Department of Physics, Chemistry, Biomedical Engineering, and Radiology, Duke University, Durham, NC, 27704

**Abstract**

Signal Amplification By Reversible Exchange (SABRE) creates "hyperpolarization" (large spin magnetization) using a transition metal catalyst and parahydrogen, addressing the sensitivity limitations of magnetic resonance. SABRE and its heteronuclear variant X-SABRE are simple, fast and general, but to date have not produced polarization levels as large as more established methods. We show here that inaccuracies in the theoretical framework for these applications, which focuses on avoided crossings (also called level anti-crossings or LACs) steer current SABRE and X-SABRE experiments away from optimal solutions. Accurate simulations show astonishingly rich and unexpected dynamics in SABRE/X-SABRE, which we explain with a combination of perturbation theory and average Hamiltonian approaches. This theoretical picture predicts simple pulse sequences with field values far from LACs (both instantaneously and on average), using different terms in the effective Hamiltonian to strategically control evolution and improve polarization transfer. Significant signal enhancements under such highly non-intuitive conditions are verified experimentally.

**Introduction**

Nuclear magnetic resonance techniques have wide applications in chemistry, physics, materials science, and medicine. Unfortunately, as the interaction of nuclei with an external magnetic field only induces small energy differences between spin states, the sensitivity of these techniques is fundamentally limited. Hyperpolarization methods artificially alter spin populations to induce polarizations far beyond thermal equilibrium by transferring polarization from a source of spin order to the target nucleus. Techniques such as dissolution Dynamic Nuclear Polarization (DNP)(*1*) and Spin Exchange Optical Pumping (SEOP)(*2*) have been in use for decades and have matured to the point of clinical applicability, but are limited by expense and need for specialized personnel to operate equipment.

An alternative method, originally introduced as Signal Amplification By Reversible Exchange (SABRE)(*3, 4*), enables large signal enhancements in many different molecules at a small fraction of the cost of other techniques. SABRE derives spin order from parahydrogen, the singlet state ($1/\sqrt{2}\,(|\alpha\beta\rangle - |\beta\alpha\rangle)$) of the hydrogen molecule, using reversible interactions between a polarization transfer catalyst (PTC), parahydrogen, and a target ligand (Figure 1). The parahydrogen creates a singlet hydride pair bound reversibly to the Ir coordination center (green), typically exchanging on and off with rates $k_H \approx 1-5\ s^{-1}$. Target ligands, such as pyridine in this example system, also reversibly bind to the complex, typically with rates $k_L \approx 5 - 50\ s^{-1}$. While the ligands are bound to the catalyst, scalar couplings between the hydrides and nuclei in the target ligand facilitate spin order transfer into bulk magnetization on the target under certain magnetic field conditions. In the decade since SABRE was introduced, many variants have been demonstrated, and several hundred different molecules have been successfully polarized.

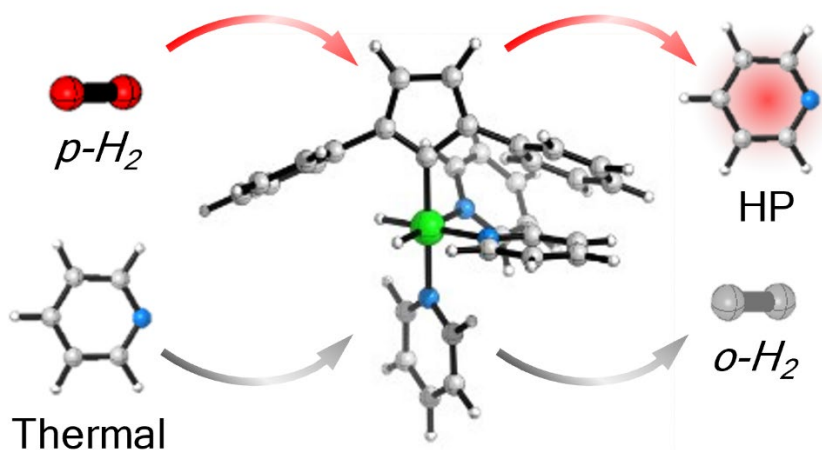

Figure 1. Exchange interactions in SABRE with an Ir(IMes) catalyst. Parahydrogen and target ligand exchange at the equatorial sites to form transient spin coupling networks for polarization transfer. The example system shown here is Ir(H)$_2$(IMes)($^{15}$N-pyr)$_3$

Hyperpolarization under SABRE is generally explained by resonant population transfer at an avoided crossing, sometimes also called a level anti-crossing (LAC). LACs provide a powerful framework in many spectroscopic applications, going back nearly a century(*5-8*). They may be rigorously used in two-level systems with a perturbation (such as an applied field that causes the system to evolve through a crossing), but their use is also ubiquitous in systems such as SABRE where the dynamics is modulated by relaxation or exchange. In the proton-only SABRE case, the large chemical shift difference between the hydride and target protons implies that a field around 6.5 mT creates a LAC. As a result, scalar couplings between the hydride and ligand efficiently connect purposefully overpopulated singlet-hydride spin states to states with net magnetization on the ligand spins(*3, 4*) Newer, extended SABRE (X-SABRE)(*9-12*) techniques have expanded the polarizable nuclear targets to include heteronuclei such as $^{15}$N, $^{13}$C, $^{19}$F, and $^{31}$P(*10, 13-16*). The states look the same as in SABRE, but now the Larmor frequency difference between $^1$H and other nuclei implies a much smaller magnetic field (~0.5 µT) to establish a matching condition where the resonance frequency difference between the hydride and target heteronucleus is comparable to the hydride-hydride scalar coupling.(*17*) In either the SABRE or X-SABRE case, as the target molecules exchange off, the solution becomes hyperpolarized, and exchange of the hydrides with fresh *p*-H$_2$ permits additional magnetization. Figure 2 shows a X-SABRE experiment where continuous application of a magnetic field B results in substantial signal enhancement.

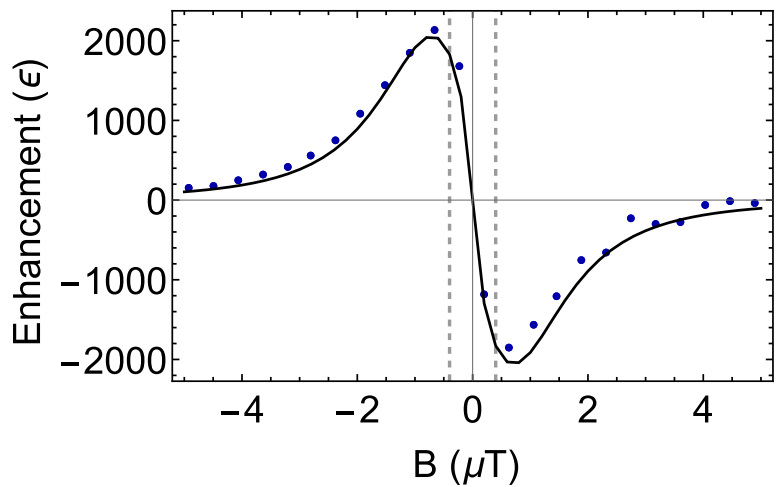

**Fig 2. X-SABRE magnetic field profile for $^{15}$N-acetonitrile.** Final polarization achieved with continuous application of a magnetic field B on an Ir(H)$_2$(IMes)(pyr)$_2$($^{15}$N-acetonitrile) PTC showing. Black line shows theoretical simulation of a Ir(H)$_2$(IMes)(pyr)$_2$($^{15}$N-acetonitrile) system ($k_N = 20\ s^{-1}$, $k_H = 2\ s^{-1}$, $[catalyst]:[ligand] = 1:20$, 30 s evolution). Data points show experimental enhancement relative to thermal magnetization measured at 1T (sample composition: 5 mM Ir(IMes), 50 mM $^{14}$N-pyridine, 100 mM $^{15}$N-acetonitrile in CD$_3$OD). Dashed lines show LAC conditions at $\pm 0.4\ \mu T$

Another attempt to use expectations derived from the LAC picture was coherently pumped SABRE SHEATH[19], which uses an optimal duration matching field at the LAC condition for polarization transfer, followed by an off-resonant field to "re-initialize" the complex under conditions where polarization transfer is halted, but exchange refreshes the ligands.

SABRE is much faster, simpler and less expensive than DNP or SEOP, and in principle should be scalable to much larger quantities. However, to date, the total achievable polarization at one time is lower than these traditional methods, which have benefitted from decades of development. In principle, with typical $^{15}$N and $^{13}$C T$_1$ values of 1 minute at low field and exchange rates $k_L \approx 5 - 50\ s^{-1}$, a single catalyst molecule could polarize thousands of targets. In practice however, the efficiency is generally a factor of 100 or so less than this. Substantial optimization is needed to improve experimentally achievable polarizations.

This paper re-examines the theoretical underpinnings of SABRE hyperpolarization to address this challenge. It is guided in part by newly developed simulation methods which account for several interacting effects all happening on the same time scale to accurately model real experimental systems.(*18*) Because this model has been shown to rigorously reproduce experimental results, it can be used to explore parameter spaces that would otherwise have gone experimentally unexplored. We show first that the LAC condition does not even give *qualitatively* correct predictions under many experimental circumstances for SABRE/X-SABRE and the coherently pumped variant. We then show that pulse sequences which move away from the LAC resonant condition (sometimes drastically and nonintuitively) can give significant enhancements.

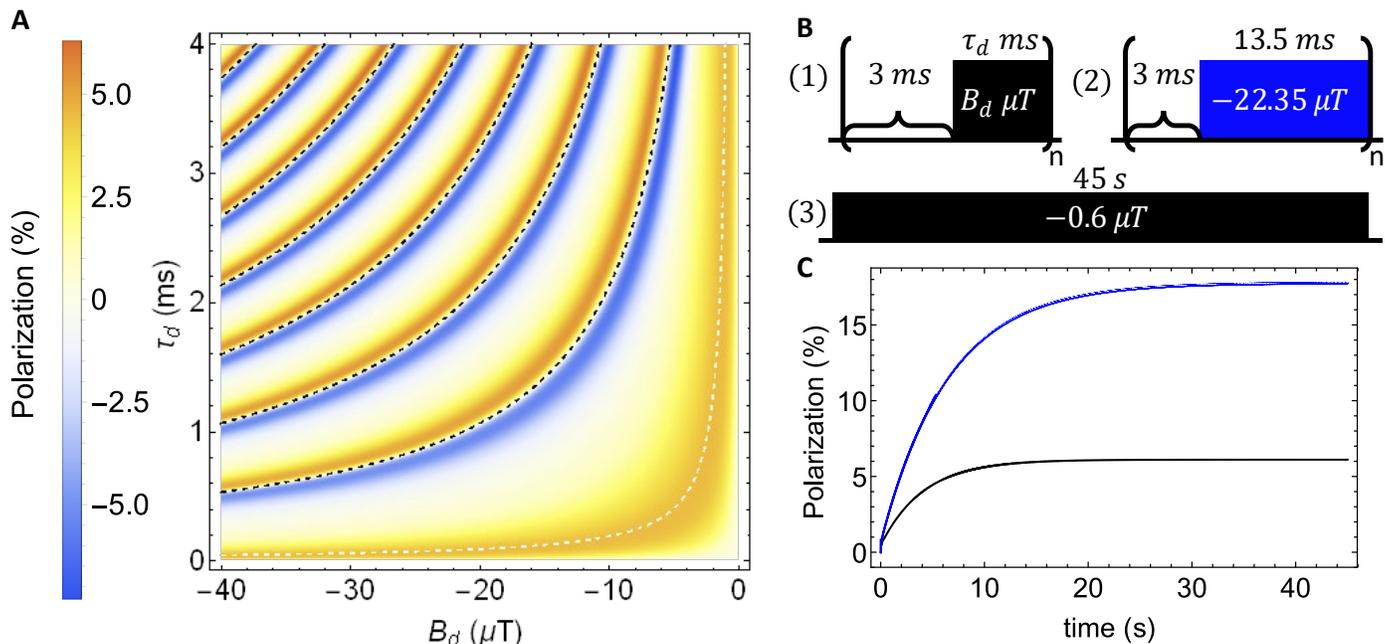

**Fig 3. Effects of several simple sequences (top right) on X-SABRE magnetization production.** (**A**) Polarization after application of pulse sequence 1 on an Ir(H)$_2$(IMes)(pyr)$_2$($^{15}$N-acetonitrile) system ($k_N = 20\ s^{-1}$, $k_H = 2\ s^{-1}$, $[catalyst]:[ligand] = 1:20$, hydride and all acetonitrile spins included). The white dashed line shows a local maximum, with an average field of $-0.6\ \mu T$, close to the LAC condition. Black dashed lines mark pulse conditions where $\theta = 2\pi B_d \tau_d (\gamma_H - \gamma_N) = 2\pi n$. The positive and negative bands around these contour lines are unexpected in the normal picture and correspond to increased magnetization. (**B**) Polarization build-up over time under sequence 2 (blue) yields a 3-fold increase in polarization over evolution under the optimal continuous field (sequence 3 in part A) shown in black.

One example to motivate this analysis is shown in Figure 3. At zero field, simple symmetry arguments discussed later imply no magnetization is ever created. A field of -22.35 µT (approaching the Earth's magnetic field) is about two orders of magnitude away from the LAC condition, so again no magnetization is produced. Yet, Figure 3 demonstrates that pulse sequences combining these field conditions can result in significant signal enhancements. Figure 3A demonstrates highly nonintuitive results with sharp features indicating magnetization enhancement in pulse sequences where the field never approaches the LAC condition transiently, or on average. On the left side of Figure 3, the white dotted line corresponds to the experimental optimum in Figure 2 (roughly at the LAC condition) on average, but all of the features at higher field are unexpected as they have no relationship to the LAC condition. The black lines correspond to combinations of field pulse areas where the precession difference between the hydride and ligand spins is an integral number of cycles ($\theta = 2\pi n$). Signal increases with larger pulse area and Figure 3C shows an example of a much larger pulse area which yields about a factor of three enhancement.

In this paper, we first show that predictions based on LACs are often incorrect even in simple cases such as continuous SABRE/X-SABRE, and even in simple spin systems. We then use a combination of average Hamiltonian theory and perturbation theory to improve the conceptual understanding of the dynamics, explain the nonintuitive structure in calculations such as Figure 3, and predict the features necessary to optimize SABRE/X-SABRE. This allows us to derive and demonstrate simple motifs, which give large signal enhancements and are robust to experimental imperfections, such as field inhomogeneity. More generally, we show that multiple fundamental limitations in polarization transfer are addressed simply by what we call two-state SABRE: two different time periods with different effective Hamiltonians can provide substantial improvements. Such sequences are easy to implement and improve the robustness of the SABRE/X-SABRE effect, particularly in the very slow and very fast exchange limits.

## Results

### Theoretical Perspectives on SABRE Dynamics

We begin by briefly reviewing the traditional theoretical framework, focusing here on two simple but illustrative cases: the three-spin case with a single target ligand nucleus $\hat{L}$ and the four-spin case with two target ligands $\hat{L}_1$ and $\hat{L}_2$. The target $\hat{L}$ is $^1$H in SABRE, and commonly $^{15}$N in X-SABRE variants such as SABRE-SHEATH. In both spin systems, two hydride spins $\hat{I}_1$ and $\hat{I}_2$ with singlet order population $S_H = (|\alpha_1\beta_2\rangle - |\beta_1\alpha_2\rangle)/\sqrt{2}$ derived from binding parahydrogen are bound to the central iridium atom, and $\alpha_n$ and $\beta_n$ are the spin-up and spin-down states of hydride spin $n$; the other three hydrogen states are $|T_{1,H}\rangle = |\alpha_1\alpha_2\rangle$, $|T_{0,H}\rangle = (|\alpha_1\beta_2\rangle + |\beta_1\alpha_2\rangle)/\sqrt{2}$ and $|T_{-1,H}\rangle = |\beta_1\beta_2\rangle$.

In the three-spin case, the ligand initially has thermally distributed populations of 50% spin up ($|\alpha_L\rangle$) and 50% spin down ($|\beta_L\rangle$). When parahydrogen and the target nucleus transiently associate with the iridium catalyst, these states are combined to give an initial density matrix of 50% $|S_H\alpha_L\rangle$ and 50% $|S_H\beta_L\rangle$. This can be written as:

$$\hat{\rho}_0 = \left(\frac{1}{4}\hat{1} - \hat{I}_1 \cdot \hat{I}_2\right) \otimes \hat{1} \quad [1]$$

It is often an excellent approximation to assume that the ligand is only coupled to one of the two hydrides (here chosen as spin 1). The evolution of these states in an externally applied magnetic field is governed by the nuclear spin Hamiltonian ($\hat{\mathcal{H}}$) shown, here in natural units ($\hbar = 1$):

$$\hat{\mathcal{H}}' = \omega_H(\hat{I}_{1z} + \hat{I}_{2z}) + \omega_L \hat{L}_z + 2\pi J_{HH}\hat{I}_1 \cdot \hat{I}_2 + 2\pi J_{HL}\hat{I}_1 \cdot \hat{L} \quad [2]$$

The Zeeman terms ($\omega_H(\hat{I}_{1z} + \hat{I}_{2z}) + \omega_L\hat{L}_z$) define the evolution of individual spins in the magnetic field ($B$) with the Larmor frequencies $\omega_n = \gamma_n B$. It is useful to write this Hamiltonian in a slightly different form:

$$\hat{\mathcal{H}}' = \omega_H(\hat{I}_{1z} + \hat{I}_{2z} + \hat{L}_z) + 2\pi J_{HH}\hat{I}_1 \cdot \hat{I}_2 + 2\pi J_{HL}\hat{I}_{1z}\hat{L}_z - (\omega_H - \omega_L)\hat{L}_z + 2\pi(\hat{I}_{1x}\hat{L}_x + \hat{I}_{1y}\hat{L}_y) \quad [3]$$

The first term, proportional to the z component of the total angular momentum, commutes with the rest of the Hamiltonian. Thus, the couplings do not *create* net z angular momentum, but they can create opposite-sign angular momentum components on the hydrogen and ligands. As an aside, this implies that SABRE evolution never creates net magnetization either, because the gyromagnetic ratios of $I$ and $L$ are the same. The created ligand magnetization is counterbalanced by ortho-hydrogen reversed magnetization which then bubbles away or relaxes quickly. X-SABRE experiments do create net magnetization because the gyromagnetic ratios of $I$ and $L$ are different.

We are ultimately interested in optimizing $\hat{L}_z$, the magnetization on the ligands, which survives dissociation of the ligand complex. This also commutes with the first term in equation [3]. It is then trivial to show this term has no effect of the production of magnetization, so we ignore it in what follows, writing

$$\hat{\mathcal{H}}' = 2\pi J_{HH}\hat{I}_1 \cdot \hat{I}_2 + 2\pi J_{HL}\hat{I}_{1z}\hat{L}_z - \Delta\omega_{HL}\hat{L}_z + 2\pi(\hat{I}_{1x}\hat{L}_x + \hat{I}_{1y}\hat{L}_y) \quad [4]$$

where the difference $\Delta\omega_{HL} = \omega_H - \omega_L$ is commonly ~20 ppm in SABRE (from chemical shift differences between protons) but $\Delta\omega_{HL} \approx 1.1\omega_H$ in SABRE-SHEATH with $^{15}$N (since $\gamma_N \approx -\gamma_H/10$). The one parameter under the control of the experimenter is $\Delta\omega_{HL}$ in the third term, which commutes with all but the last term in equation [3].

The extremely simple expression in equation [4] is powerful for providing physical insight. We start by looking at the traditional, continuous field SABRE/X-SABRE case. The LAC condition is at fields where the overpopulated $|S_H\alpha_L\rangle$ state and empty $|T_{1,H}\beta_L\rangle$ state have nearly diagonal elements and the coupling mixes the two states. One 3x3 block matrix is shown below, the rest are written out in the Supplementary Information.

$$\widehat{\mathcal{H}}' = \begin{array}{c} \\ \langle S_H\alpha_L| \\ \langle T_{1,H}\beta_L| \\ \langle T_{0,H}\alpha_L| \end{array} \begin{pmatrix} |S_H\alpha_L\rangle & |T_{1,H}\beta_L\rangle & |T_{0,H}\alpha_L\rangle \\ -2\pi J_{HH} & \dfrac{\pi J_{HL}}{\sqrt{2}} & \dfrac{-\pi J_{HL}}{2} \\ \dfrac{\pi J_{HL}}{\sqrt{2}} & \dfrac{\pi(J_{HH}-J_{HL})}{2}+\Delta\omega_{HL} & \dfrac{\pi J_{HL}}{\sqrt{2}} \\ \dfrac{-\pi J_{HL}}{2} & \dfrac{\pi J_{HL}}{\sqrt{2}} & 0 \end{pmatrix} \quad [5]$$

The diagonal elements for $|S_H\alpha_L\rangle$ and $|T_{1,H}\beta_L\rangle$ are equal when $\Delta\omega_{HL} = -2\pi J_{HH} + \pi J_{HL}/2$. Typically, in SABRE-SHEATH where the nitrogen nucleus is bound directly to the Ir catalyst, we find $J_{HH} \approx -8\ Hz$ and $J_{HL} \approx -20\ Hz$, which corresponds to $B \approx -0.4\ \mu T$, consistent with Figure 2. In SABRE (with a proton target) the hydride and target nuclei are separated by four or more bonds, giving a smaller $J_{HL} \approx 1\ Hz$ and $\Delta\omega_{HL} \approx -2\pi J_{HH}$ matching condition. Common chemical shift differences between target and hydride protons indicate a LAC condition around 6.5 mT.

Another LAC is found at positive magnetic field when $\Delta\omega_{HL} = 2\pi J_{HH} - \pi J_{HL}/2$, now between the $|S_H\beta_L\rangle$ state and $|T_{-1,H}\alpha_L\rangle$ states, which would create opposite sign magnetization on the ligands. This is the basic approach behind the design of essentially all low-field SABRE and X-SABRE experiments. Approximating $|S_H\alpha_L\rangle$ and $|T_{1,H}\beta_L\rangle$ as a two-level system leads, for example, to the coherently pumped SABRE idea that the population can oscillate back and forth only when the diagonal elements are nearly matched, and thus interrupting the evolution by switching to high field at a maximum can give improved polarization.(*19*)

A similar calculation for the four-level system (in the Supplementary Information) gives a 4x4 matrix and two 2x2 matrices where the ligand-hydride coupling can create ligand magnetization. The states $|S_H S_L\rangle$ and $|T_{1,H}T_{-1,L}\rangle$ have the same diagonal values when $\Delta\omega_{HL} = -2\pi(J_{HH} + J_{LL}) + \pi J_{HL}$. Additionally, the states $|S_H T_{1,L}\rangle$ and $|T_{1,L}S_L\rangle$ have the same diagonal values when $\Delta\omega_{HL} = -2\pi(J_{HH} - J_{LL})$. Both of these cases create net spin-down magnetization on the ligand nucleus at negative fields. Parallel arguments apply to the two corresponding LACs at positive field, which create positive magnetization.

The point of the approximation of these cases as two-level systems is to give an intuition for the optimal field in SABRE polarization transfer. Of course, reducing a complex quantum mechanical system to this extent is an oversimplification, which is made even worse when there are multiple couplings. In the X-SABRE case, $J_{HL}$ often exceeds $J_{HH}$, which couples in states other than the initial and target states (for example, $|T_{0,H}\alpha_L\rangle$ in equation [5]). In addition, the catalyst dynamics is exceedingly complex, as exchange rates (both of the ligands and of the parahydrogen), internal couplings, and resonance frequency differences are all very similar. Still, the commonly explored cases tend to look like Figure 2, in broad agreement with the LAC prediction.

We have recently developed a new numerical modeling approach for exchanging systems, which is extremely useful in the practical limits commonly seen in SABRE.(*18*) These simulation methods handle exchange to infinite order in perturbation theory, making it feasible to take large time steps in the spin evolution without compromising accuracy. These simulations have already shown robust agreement with experimental results. This, in turn, lets us explore a wider range of conditions than is easily accessible experimentally and is shown here to be directly validated by targeted experiments. We start by looking at various conditions in standard continuous field SABRE and X-SABRE (Figure 4) to show a surprising result: *except under specific exchange regimes, the LAC picture is a poor oversimplification*, and the optimal conditions are a very strong function of exchange rate (which never appears in the LAC calculation). Remarkably, the curves look quite similar in the three and four-spin models (in the Supplementary Material), and even in more complex systems.

For very slow exchange, where one might expect the LAC prediction to perform best, the optimum field is significantly lower than the LAC prediction and decreases instead of increasing for larger $J_{HL}$. For very rapid

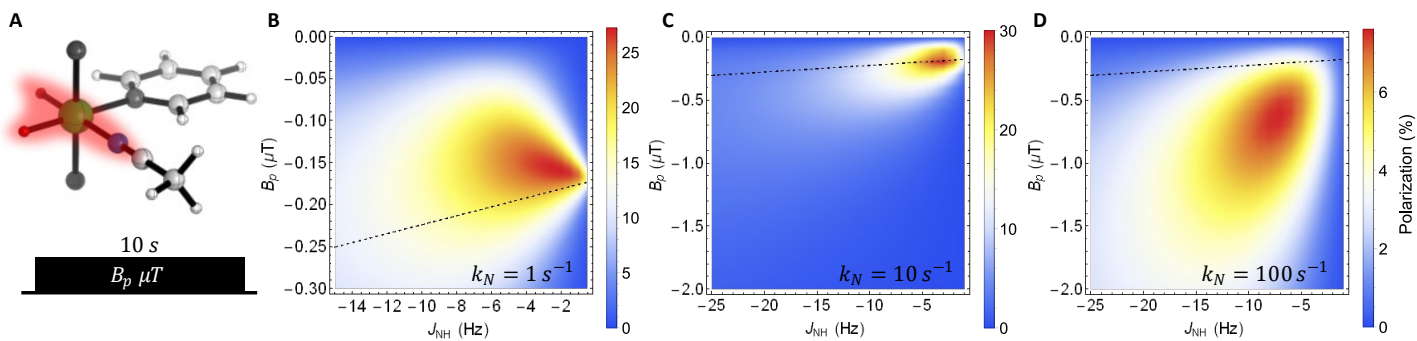

**Fig 4. Discrepancies between optimal polarization transfer fields and LAC matching conditions in continuous field experiments.** (**A**) Truncated AA'X spin system with incorporated spins highlighted in red and the pulse sequence used in the simulations. (**B**) Polarization after application of the pulse sequence in A under a variety of field conditions $B_p$ and $J_{NH}$ couplings and $k_L = 1\ s^{-1}$, (**C**) $k_L = 10\ s^{-1}$, and (**D**) $k_L = 100\ s^{-1}$. Black dashed lines correspond to the AA'X LAC condition $\Delta\omega_{HL} = 2\pi J_{HH} - \pi J_{NH}/2$. Additional simulation parameters: $[catalyst]:[ligand] = 1:20, k_H = 1\ s^{-1}$.

exchange, the optimum field far exceeds the LAC prediction. Thus, much of the experimental difficulty in getting good polarization outside of a narrow exchange rate range can be attributed to the need to change the field.(*17*)

Figure 4 also shows that X-SABRE commonly uses a larger than ideal $J_{HL} \approx -20\ Hz$. Figure 4 would look identical for SABRE, with a rescaling of the y axis by about 30,000 (the dotted line would be at $B_d = 6.5\ mT$ for $J_{HH} = 0\ Hz$), which implies that normal SABRE couplings ($\approx 1\ Hz$) are significantly smaller than optimal except for very slow exchange.

For coherently pumped SABRE (Figure 5) at the LAC condition, the off-diagonal matrix elements in equation [5] are expected to pump magnetization in and out of the ligand, with an oscillation frequency of $J_{HL}/\sqrt{2}$ (hence a maximum at $\tau_p = 1/(\sqrt{2}J_{HL})$). Coherently pumped SABRE SHEATH pulse sequences use this understanding and would be expected to produce the most efficient transfer (in a two-level approximation) by going to the LAC condition. Again, this result is only correct for a very limited set of exchange rates. For rapid exchange in particular, using the LAC condition leads to significant signal attenuation. Again, larger spin systems (in the Supplementary Information) provide very similar results.

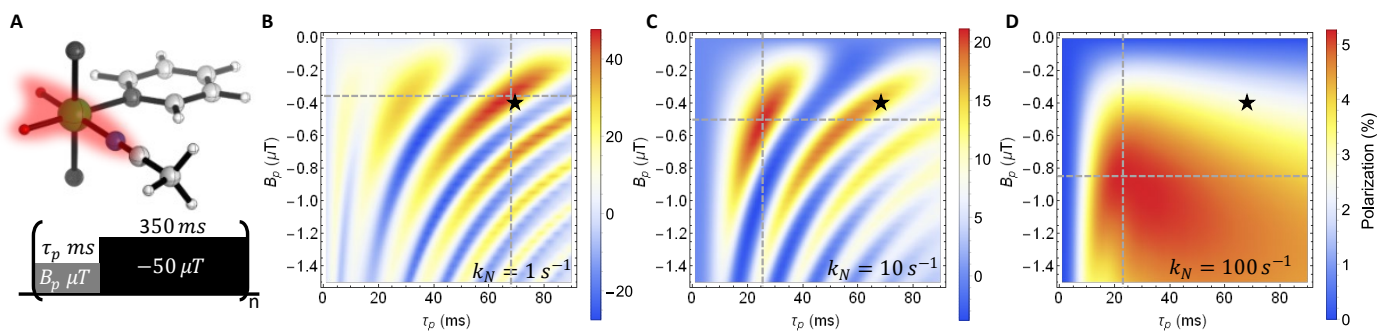

**Fig 5. Discrepancies between optimal polarization transfer fields and LAC matching conditions in coherently pumped SABRE SHEATH experiments.** (**A**) Truncated AA'X spin system with incorporated spins highlighted in red and the pulse sequence used in the simulations. (**B**) Polarization after application of the pulse sequence in A for 30 seconds under a variety of evolution field conditions $B_p$ and durations $\tau_p$ and $k_L = 1\ s^{-1}$, (**C**) $k_L = 10\ s^{-1}$, and (**D**) $k_L = 100\ s^{-1}$. Dashed gray lines indicate maximum polarization conditions; the star marks the expected LAC matching condition. Additional simulation parameters: $[catalyst]:[ligand] = 1:5, k_H = 1\ s^{-1}$.

Figures 4 and 5 show that, even for the simplest strategies for generating SABRE in common use, the LAC approach has limited predictive value. With this understanding of the flawed and often misleading approximations previously used to guide SABRE experiments, we will now propose a different strategy to understand SABRE and X-SABRE evolution which will account for the unexpected results shown in Figure 3 and provide intuitive justifications for new pulse sequences.

Evolution of the density matrix $\hat{\rho}$ under the nuclear spin Hamiltonian is dictated by the Liouville von Neumann equation $\partial_t \hat{\rho} = i[\hat{\rho}, \hat{\mathcal{H}}]$. Using a Taylor series expansion of the solution of this differential equation, we can separate out the time dependence for the generation of various spin states:

$$\hat{\rho}(t) = \hat{\rho}_0 + \frac{(it)}{1!}[\hat{\rho}_0, \hat{\mathcal{H}}] + \frac{(it)^2}{2!}\left[[\hat{\rho}_0, \hat{\mathcal{H}}], \hat{\mathcal{H}}\right] + \frac{(it)^3}{3!}\left[\left[[\hat{\rho}_0, \hat{\mathcal{H}}], \hat{\mathcal{H}}\right], \hat{\mathcal{H}}\right]$$
$$+ \frac{(it)^4}{4!}\left[\left[\left[[\hat{\rho}_0, \hat{\mathcal{H}}], \hat{\mathcal{H}}\right], \hat{\mathcal{H}}\right], \hat{\mathcal{H}}\right] + \cdots \qquad [6]$$
$$= \hat{\rho}_0 + it[\hat{\rho}_0, \hat{\mathcal{H}}] + \frac{it^2}{2!}[\dot{\rho}, \hat{\mathcal{H}}] + \frac{it^3}{3!}[\ddot{\rho}, \hat{\mathcal{H}}] + \frac{it^4}{4!}[\dddot{\rho}, \hat{\mathcal{H}}] + \cdots$$

The initial state for this system consists of singlet order on the hydride nuclei and thermal polarization on the target nucleus to give the initial density matrix in equation [1]. For example, starting with the first derivative term in the series, $(it[\hat{\rho}_0, \hat{\mathcal{H}}])$, $\hat{\rho}_o$ commutes with everything except the $J_{HL}$ term in the Hamiltonian, giving

$$\dot{\rho}t = it[\hat{\rho}_o, \hat{H}] = it[\hat{I}_1 \cdot \hat{I}_2, 2\pi J_{HL}\hat{I}_1 \cdot \hat{L}]$$
$$= 2\pi J_{HL} t(-\hat{I}_{1z}\hat{I}_{2x}\hat{L}_y + \hat{I}_{1y}\hat{I}_{2x}\hat{L}_z + \hat{I}_{1z}\hat{I}_{2y}\hat{L}_x - \hat{I}_{1x}\hat{I}_{2y}\hat{L}_z - \hat{I}_{1y}\hat{I}_{2z}\hat{L}_x + \hat{I}_{1x}\hat{I}_{2z}\hat{L}_y) \qquad [7]$$

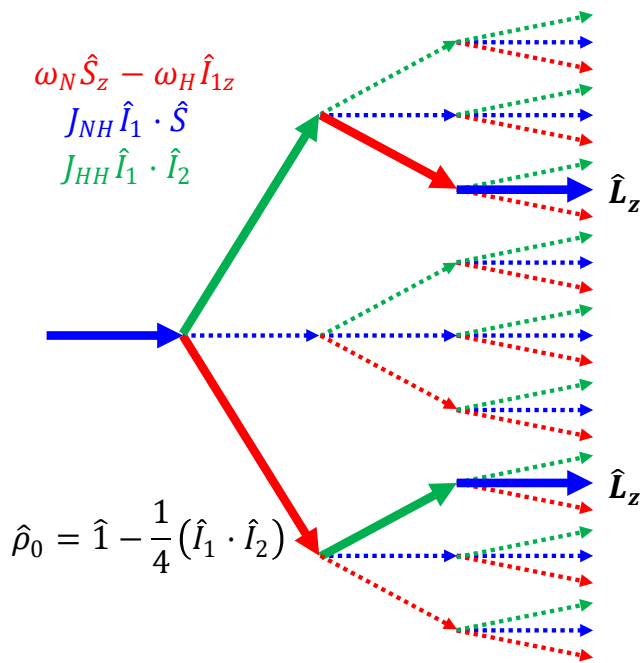

**Fig 6. Demonstration of coherence pathways.** Evolution of the initial SABRE density matrix, $\hat{\rho}_0$, under each of the terms in the nuclear spin Hamiltonian. The polarization transfer pathway is highlighted in bold and off-target pathways are shown with dashed lines.

Creation of $\hat{L}_z$ order from equation [1], using the Hamiltonian in equation [3] and the expansion in equation [6], can easily be shown to require at least four commutators. Commutation of a bilinear operator with any $n$-spin direct product operator always adds or subtracts one spin, so the first derivative in equation [7] will require two more commutators with bilinear operators to transform the initial spin order into $\hat{L}_z$. One of those must be from the $J_{HH}$ terms as it is the only bilinear operator with spin 2, and one must be from $J_{HL}$. In addition, only the Zeeman term $\Delta\omega_{HL}$ distinguishes between spin up and spin down (by symmetry no magnetization is ever formed at $B = 0 \mu T$), and this term must be involved in the commutator pathway as well. Thus, the functional dependence of the leading term must be $J_{HL}^2(\Delta\omega_{HL})J_{HH}$ in the fourth derivative leading to magnetization build-up initially proportional to $t^4$. A full evaluation (in the Supplementary Information) shows:

$$\dddot{\rho}\frac{t^4}{4!} = (2\pi)^3 J_{HL}^2(\Delta\omega_{HL})J_{HH}\frac{t^4}{4!}(2\hat{L}_z - 2\hat{I}_z) \qquad [8]$$

The $J_{LH}$ term (specifically, the non-secular component of the $J_{LH}$ term ($\hat{I}_{1x}\hat{L}_x + \hat{I}_{1y}\hat{L}_y$)) is needed in the first and fourth commutators. The $J_{HH}$ and $\Delta\omega_{HL}$ terms are needed for the second and third commutators, but the order is unimportant as the terms commute with each other. Figure 6 illustrates the "flow" into operator space in terms of the three critical spin interactions. The goal is to maximize the two pathways which create magnetization, while minimizing the off-target pathways.

This approach helps explain the significant deviations from the LAC condition for high and low exchange rates shown in Figures 4 and 5. In the rapid exchange limit, the achievable magnetization per unit time is expected to scale as $k^{-3}$, because doubling $k$ (halving the lifetime of the complex) reduces the term in equation [8] by 16, but gives twice as many molecules which can be (partially) polarized per unit time. In this case, higher values of the applied field (to achieve significant phase shift over the complex lifetime) are strongly favored, which is again consistent with Figures 4-5.

In the limit of low exchange, the optimal strategy is to maximize the ligand magnetization at around the average exchange time. The applied magnetic field is best understood as generating a phase shift; for example, as shown in the Supplementary Information, it converts terms such as $\hat{I}_{1x}\hat{L}_x$ into $\hat{I}_{1x}\hat{L}_y$, which can be converted into $\hat{L}_z$ by the $J_{HL}$ coupling. The optimal way to do this, without over-pumping other operators, is to force this term produce about $\pi/2$ of phase shift over the lifetime of the complex. This implies *the optimum field is strongly dependent on the exchange rate*. This explains Figure 4b, where the optimum field decreases with increasing $J_{HL}$; the term in equation [8] is kept constant by reducing $\Delta\omega_{HL}$ when $J_{HL}$ is increased, just the opposite of what is expected by the LAC condition.

The significance of this rotation can be seen in Figure 7, which explicitly shows the effect of varying $\Delta\omega_{HL}$ for a zero field-high field sequence (here and in all figures except Figures 4 and 5, we include all the spins in the hydride and the ligand). With the pulse sequence in Figure 7A, the spin system evolves under $\hat{I}_x\hat{L}_x + \hat{I}_y\hat{L}_y$ during the zero-field pulse and then acquires some phase shift $\theta$ under the $\Delta\omega_{NH}$ term during the high-field pulse. In this calculation the exchange rate is $20\ s^{-1}$ giving a mean lifetime before exchange of $\sim 35\ ms$. When $\tau_p = 3\ ms$ as in the blue curve in Figure 7B, the zero-field/high-field alternation is seen about ten times by the average complex, and efficient transfer is done with a small phase increment in each cycle (hence the maxima occur with small deviation from $2\pi n$). This also explains why the pulse sequence explored in Figure 3 showed such sharp features at small rotations away from integral numbers of cycles about $\Delta\omega_{NH}$ - for the short $\tau_d$ and $\tau_p$ values shown there, the complex sees many pulse repetitions during its lifetime. In contrast, when $\tau_p = 50\ ms$ in Figure 7, most of the complexes see at best a single high field pulse, the process is much less efficient, and a single pulse needs to produce a substantial phase shift.

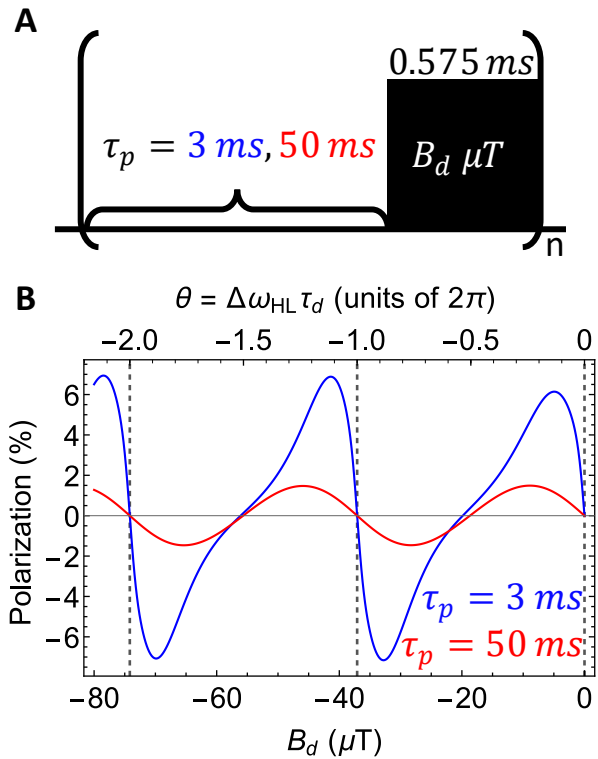

Fig 7. Flip angle and PTC lifetime interactions. (A) Pulse sequence used in simulations. (B) Final polarization after application of the sequence in A for values for $\tau_p = 3\ ms$ (blue) and $\tau_p = 50\ ms$ (red) with $B_p = 0\ \mu T$. Vertical dashed lines show $2\pi n$ rotations. Simulation parameters: Ir(H)$_2$(IMes)(pyr)$_2$($^{15}$N-acetonitrile) spin system, [catalyst]: [ligand] = 1: 20, $k_L = 20\ s^{-1}$, $k_H = 2\ s^{-1}$, experiment duration 5 seconds

We may further understand the nonintuitive dynamics presented here using average Hamiltonian theory. For the Hamiltonian in equation [3], only one parameter ($\Delta\omega_{HL}$) is directly experimentally accessible. If we use this parameter to create a toggling frame we get:

$$\tilde{\mathcal{H}}'(t) = U\hat{\mathcal{H}}'(t)U^\dagger; U(t) = \exp(-i\Delta\omega_{HL}\hat{L}_z t)$$
$$\tilde{\mathcal{H}}'(t) = 2\pi J_{HH}\hat{I}_1 \cdot \hat{I}_2 + 2\pi J_{HL}\hat{I}_{1z}\hat{L}_z + 2\pi J_{HL}(M(t)(\hat{I}_{1x}\hat{L}_x + \hat{I}_{1y}\hat{L}_y) + N(t)(\hat{I}_{1y}\hat{L}_x - \hat{I}_{1x}\hat{L}_y)) \quad [9]$$

where $M(t) = \cos(\Delta\omega_{HL}t)$, and $N(t) = \sin(\Delta\omega_{HL}t)$. In the constant field case (fixed $\Delta\omega_{HL}$, duration $T$), the integrals are easily evaluated, and we can immediately write the average Hamiltonian (the lowest order term in the Magnus expansion) as

$$\tilde{\mathcal{H}}^{(0)}(T) = U^\dagger(T)\left(\frac{1}{T}\int_0^T \tilde{\mathcal{H}}'(t)\,dt\right)U(T) \quad [10]$$

$$\tilde{\mathcal{H}}^{(0)}(T) = U^\dagger(T)\left(2\pi J_{HH}\hat{I}_1 \cdot \hat{I}_2 + 2\pi J_{HL}\hat{I}_{1z}\hat{L}_z \right.$$
$$\left. + 2\pi J_{HL}\left(\frac{\sin(\Delta\omega_{HL}T)}{\Delta\omega_{HL}T}(\hat{I}_{1x}\hat{L}_x + \hat{I}_{1y}\hat{L}_y) + \frac{1-\cos(\Delta\omega_{HL}T)}{\Delta\omega_{HL}T}(\hat{I}_{1y}\hat{L}_x - \hat{I}_{1x}\hat{L}_y)\right)\right)U(T)$$

In the most common applications of average Hamiltonian theory, the sequence is cyclic making $U(T) = \hat{1}$. Here we will not impose that restriction; in equation [10], $U(T) = \exp(-i\Delta\omega_{HL}T\hat{L}_z)$. This will turn out to be important when we use two different fields in the pulse sequence. At all times, $U(T)$ commutes with both the initial density matrix and the desired final state.

There are fundamental differences between the Hamiltonians in equations [3] and [10]. The most obvious of which are the reduction of the nonsecular component of the H-L coupling, and the introduction of an operator of the form $(\hat{I}_{1y}\hat{L}_x - \hat{I}_{1x}\hat{L}_y)$. This connects the same states as the normal non-secular term, but with a $\pi/2$ phase shift for the off-diagonal operators.

Equation [12] quantifies the extent to which a resonance frequency difference truncates the nonsecular components of the scalar coupling, and it can be used in two fundamentally different ways. One way (the dominant effect in Figure 3) is to use this to decrease the average value of the effective coupling. For example, for the -22.35 µT case illustrated in Figure 3, the coupling is nearly fully truncated during the 13.5 ms pulse, so the average coupling over the entire zero field/high field cycle is only about one-fifth of the normal coupling (which, as Figure 4 shows, improves the efficiency). This also explains the asymmetry in Figures 3 and 7 between $2\pi n$ and $(2n + 1)\pi$ rotations: in the former case, the scalar coupling during the high field pulse is completely truncated by equation [12], but it is not in the latter case, and this is more consequential for shorter values of $\tau_p$.

With two nonzero fields, a different effect becomes important: the two intervals can produce different effective Hamiltonians. For example, one can adjust the rotation angles of the two blocks to produce different combinations of $(\hat{I}_{1x}\hat{L}_x + \hat{I}_{1y}\hat{L}_y)$ and $(\hat{I}_{1y}\hat{L}_x - \hat{I}_{1x}\hat{L}_y)$. The combination of operators $(\hat{I}_1 \cdot \hat{I}_2)$, $(\hat{I}_{1x}\hat{L}_x + \hat{I}_{1y}\hat{L}_y)$, and $(\hat{I}_{1y}\hat{L}_x - \hat{I}_{1x}\hat{L}_y)$ can then convert the initial density matrix into $\hat{L}_z$ magnetization in the *third* commutator, rather than the fourth as in equation [8]. Again, the optimum occurs for rotations $\theta \approx 2\pi$ or greater, because the overall magnitude of the residual couplings is decreased.

**Optimizing Non-resonant Pulse Sequences for Polarization Transfer**

In this section, we use the concepts just developed to produce optimized pulse sequences for X-SABRE polarization transfer and verify the results experimentally. We start by noting that the restriction to zero field in Figure 3 for one of the intervals is not necessary; the structure merely shifts (Figure 8). To test the simulations experimentally, a SABRE experiment was prepared on solutions with 100 mM of $^{15}$N-acetonitrile, 33 mM natural abundance pyridine, and 5 mM IrIMes(COD)Cl (IMes = 1,3-bis(2,4,6-trimethylphenyl)-imidazol-2-ylidene, COD = 1,5-cyclooctadiene) in methanol-d$_4$. The solution was bubbled with 43% parahydrogen gas to allow for the formation of the active PTC (Ir(H)$_2$(IMes)(pyr)$_2$($^{15}$N-acetonitrile)). After 30 minutes of continuous bubbling at lab field (~2.6 G), the sample was transferred into a triple µ-metal shield with a compensating solenoid powered by a standard function generator for hyperpolarization under the various two-state SABRE pulse sequences for a total duration of 60 s. After hyperpolarization, the sample is transferred into a 1T $^{15}$N Magritek NMR for signal detection.

We varied either the high field strength ($-22\ \mu T < B_d < 0\ \mu T$, $\tau_d = 2\ ms$), or duration ($0\ ms < \tau_d < 3\ ms$, $B_d = -34\ \mu T$) for the high-field pulse in a two-state SABRE sequence while maintaining a constant low field pulse ($B_p = -0.5\ \mu T$, $\tau_p = 3\ ms$). Under these conditions, we clearly see the final polarization levels in Figure 8 oscillating at a frequency corresponding to $\Delta\omega_{HL}$. In agreement with the theoretical predictions, maximum positive and negative signal enhancements for this set of pulse parameters occurs at flip angles of $\theta = 2\pi n \pm \pi/6$ with an 9% improvement in the total signal enhancement over a continuous field experiment at $B = -0.5 \, \mu T$. We demonstrate that separation of the evolution under different terms in the Hamiltonian is experimentally quite simple and improves the achievable polarization.

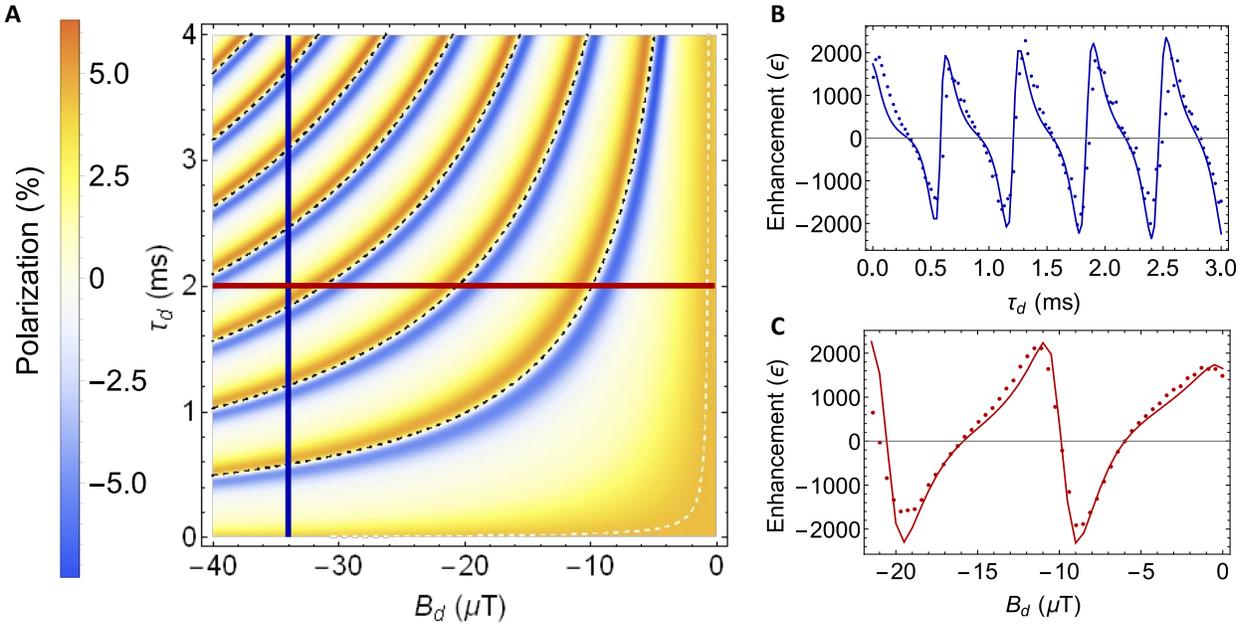

**Fig 8. Experimental validation of theoretical predictions. (A)** Theoretical effects of a pulse sequence similar to Figure 3, except the zero-field portion is replaced by a field near the LAC condition (0.5 mT). The white curve again traces out the local maximum at the lowest field. **(B)** Comparison of theoretical calculations with experimental data holding $B_d = -34 \, \mu T$ and varying $\tau_d$ along the blue line in A. **(C)** Comparison of theoretical calculations with experimental data holding $\tau_d = 2 \, ms$ and varying $B_d$ along the red line in A. Simulation parameters: $k_L = 20 \, s^{-1}$, $k_H = 2 \, s^{-1}$, $[catalyst]:[ligand] = 1:20$, experiment duration 5 seconds.

In the regime of Figure 8, where $\theta$ is small and $J_{HL}$ is large, the signal enhancements are modest. More generally, it is clear from Figure 4 that X-SABRE is fundamentally limited because $J_{HL} \gg J_{HH}$. So, a general strategy for signal enhancement uses field blocks which produce a highly reduced average coupling, as well as the desired phase shift. For example, consider the sequence in Figure 9 below. We fix one block at a field strength $B_d = -34 \, \mu T$ and duration $\tau_d = 0.575 \, ms$ (corresponding to a total rotation of $\theta = -2\pi + \pi/6$). From equation (12), this reduces the average coupling during that period by about a factor of 7. With the lower field pulse duration fixed at $\tau_p = 3 \, ms$, the field strength was swept through $-18 \, \mu T < B_p < 18 \, \mu T$. Once again, there is excellent agreement between theory and experimental results. We see the predicted large sidebands with significant signal enhancements at field strengths resulting in total rotations of $2\pi n \pm \epsilon$ about the resonance offset term yielding a maximum enhancement of about a factor of 3.

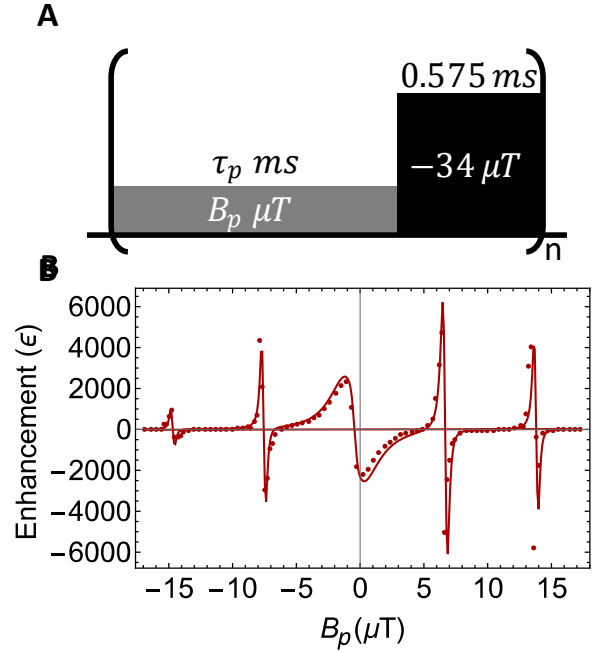

**Fig 9. Non-zero field sweep.** Experimental results (points) and numerical simulation (solid line) for the two-field SABRE sequence shown with various lower field pulse magnitudes. Simulation parameters: Ir(H)$_2$(IMes)(pyr)$_2$($^{15}$N-acetonitrile) spin system, $k_L = 20 \, s^{-1}$, $k_H = 2 \, s^{-1}$, $[catalyst]:[ligand] = 1:20$, 5 sec experimental duration.

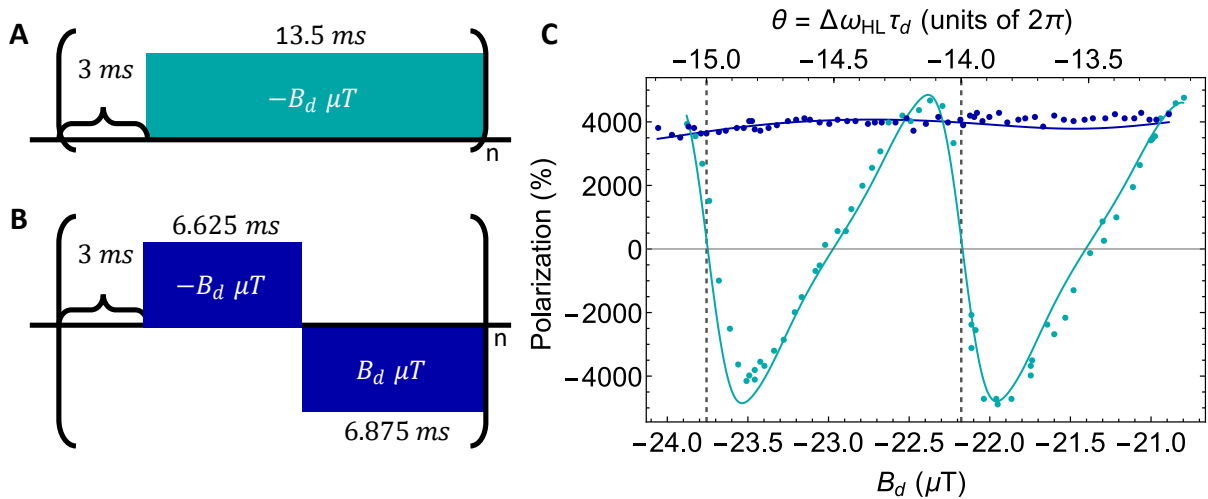

**Fig 10. Compensated and uncompensated zero-field pulse sequences**. (**A**) 2 state pulse sequence with long higher field pulse duration ($\tau_d = 13.5\ ms$) resulting in $B_d\tau_d(\gamma_H - \gamma_N) \approx 27\pi - 29\pi$ rotation across the measured fields. (**B**) Inhomogeneity compensated 2 state pulse sequence with a bipolar higher field pulse causing a net rotation of $\sim\pi/2$. (**C**) Theoretical and experimental results for the uncompensated (cyan) and compensated (blue) 2 state pulse sequences. Experiment run for 60s on a sample composition of 5 mM Ir(IMes), 50 mM $^{14}$N-pyridine, 100 mM $^{15}$N-acetonitrile in CD$_3$OD. Simulation parameters: Ir(H$_2$)(IMes)(pyr)$_2$($^{15}$N-acetonitrile) spin system, $[catalyst]:[ligand] = 1:20$, $k_L = 20\ s^{-1}$, $k_H = 2\ s^{-1}$, experiment duration 15 seconds.

## Compensation for Experimental Inhomogeneity

The sideband resonances offer further experimental benefits in addition to increased overall polarization. The zeroes come from a lower field pulse of flip angle $\theta \approx 2\pi n - \pi/5$. The case $n = 1$ corresponds to the resonance at 6.5 $\mu T$. All but the near-zero field maxima and minima are offset from exact $2\pi(n-1)$ rotations by about $\pi/30$. In the event of moderate inhomogeneities which alter the applied field by some coefficient, the net flip angle for the $n = 1$ resonance would remain almost unchanged. Two-state pulse sequences with these parameters increase the achievable polarization and are insensitive to field inhomogeneities.

Based on the theoretical picture presented so far, it is simple to modify sequences to compensate for significant inhomogeneity. Consider, for example, the zero-field/high-field sequence of Figure 3, where substantial signal gains require very long evolution times. This pulse needs to provide a small net phase shift, so the signal is highly sensitive to the exact amplitude (Figure 10) and even a small amount of inhomogeneity will change this rotation enough to randomize the phase shift required in the evolution. However, a simple modification, replacing the long -22.35 µT pulse with a slightly offset or slightly asymmetric square wave alternating between +22.35 µT and -22.35 µT, almost completely eliminates this sensitivity. The slight asymmetry in this sequence is important. For a symmetric square wave of any amplitude, the net area is zero, $U(T) = \hat{1}$ in equation [10], and the necessary phase shift does not accumulate between pulse repetitions. However, the pulse sequence in Figure 10B has a $\pi/2$ net rotation under the alternating high field pulse to gain the necessary phase shift

## Discussion

Traditional SABRE polarization strategies have focused on sequences which establish resonance between spin states to pass population between states (the avoided crossing or level anti-crossing (LAC) condition). The theoretical justification for this is obviously an approximation, and all approximations have limits. We show here, though, that sequences where neither the instantaneous nor average field ever approach a LAC condition can enhance polarization transfer and that the optimal field conditions vary greatly with the exchange rate. Strategic manipulation of individual terms in the Hamiltonian increases the rate of polarization transfer by favoring the interaction pathways which lead to magnetization. Because this transfer is largely dependent on the resonance frequency difference between the singlet hydrides and the target nucleus, it is possible to separately target different nuclei in the same compound and it would be reasonable to extend this technique into larger field regimes like that of SABRE. Using a new approach to understanding the evolution of SABRE systems sequentially through multiple interactions with the Hamiltonian, we gain a deeper understanding of the system dynamics, increase the total achievable polarization for experiments, and open up the potential to target different nuclei with a high specificity.

Our theoretical results show that at least in this system, where exchange and coherent evolution occur on comparable timescales, it is more accurate to think of the resonance frequency difference as *generating phase shifts* rather than avoided crossings. The structural features in Figures 3 and 7, for example, have no relationship to frequency matching, but phase matching in the evolution is central. Thus, for very high exchange rates, it is critical to go to much larger magnetic fields than the LAC conditions predict; you want a significant phase shift during the complex lifetime. What we have presented is relevant for both SABRE and X-SABRE, but the pulse sequence manipulations we propose are certainly easier for X-SABRE, where rapid magnetic field shifts are trivial and the ligand-hydride coupling is larger than its optimum.

The apparatus we use here is designed to have extremely good field homogeneity, but as shown in Figure 10 that is not necessary with compensated sequences, which would allow for the use of larger samples where inhomogeneities would be more prevalent. We believe the optimal strategy is to create polarization on the ligand only over a time comparable to the ligand lifetime. Within those constraints, many sequences are possible by the guidelines presented here. For example, sequences which omit the zero-field time, and simply use a slightly asymmetric square wave timed to reduce $J_{HL}$ significantly, perform about as well; they have another potential advantage, in that $T_1$ relaxation times are generally somewhat longer at tens of microtesla than they are at extremely low fields.(*20*) In fact, it is completely trivial experimentally to generate arbitrary waveforms on these timescales and with these field strengths. The basic approach presented here opens up a wide new range of sequences which promise to significantly improve both the generality and effectiveness of SABRE, thus enabling new applications.

**Acknowledgments**

**Funding:**
This work was supported by the National Science Foundation under grant CHE-2003109.

**Competing interests:** Authors declare that they have no competing interests.

**Data and materials availability:** All data are available in the main text or the supplementary materials. Additional data and materials related to this paper may be requested from the authors.

# Improving SABRE hyperpolarization with highly non-intuitive pulse sequences: moving beyond avoided crossings to describe dynamics

Shannon L. Eriksson,[1,2] Jacob R. Lindale,[1] Xiaoqing Li[3], Warren S. Warren,[1,2,3,4*]

[1]Department of Chemistry, Duke University, Durham, NC, 27708
[2]School of Medicine, Duke University, Durham, NC, 27708
[3]Department of Physics, Duke University, Durham, NC, 27708
[4]Department of Physics, Chemistry, Biomedical Engineering, and Radiology, Duke University, Durham, NC, 27704

## Supplementary Text

Representative Hamiltonians

SABRE spin networks can take one of 2 primary conformations. The first is when a coligand is present in solution and the PTC binds to one target ligand and one coligand to form an AA'B spin system. Additional spins on the target molecule can couple in to the network, but the simplest model for this system can get truncated at 3 spins. The second is when the target ligand is the only other compound in solution and the PTC binds to two targets to form an AA'BB' spin system. Again, this is the simplest truncation of many experimentally relevant systems.

The nuclear spin Hamiltonian for a SABRE AA'B spin system with an arbitrary ligand L breaks out into 4 subspaces in a basis set with singlet-triplet on the hydrides and Zeeman on the target nucleus, two 1x1 and two 3x3, shown below:

$$\langle T_{1,H}\alpha_L| \quad \left(\frac{2\pi J_{HL}}{4} + \omega_H\right)$$

with basis $|T_{1,H}\alpha_L\rangle$.

$$\begin{array}{c|ccc}
 & |S_H\alpha_L\rangle & |T_{1,H}\beta_L\rangle & |T_{0,H}\alpha_L\rangle \\
\hline
\langle S_H\alpha_L| & -2\pi J_{HH} & \frac{\pi J_{HL}}{\sqrt{2}} & \frac{-\pi J_{HL}}{2} \\
\langle T_{1,H}\beta_L| & \frac{\pi J_{HL}}{\sqrt{2}} & \frac{\pi(J_{HH} - J_{HL})}{2} + \omega_H - \omega_L & \frac{\pi J_{HL}}{\sqrt{2}} \\
\langle T_{0,H}\alpha_L| & \frac{-\pi J_{HL}}{2} & \frac{\pi J_{HL}}{\sqrt{2}} & 0
\end{array}$$

$$\begin{array}{c|ccc}
 & |T_{0,H}\beta_L\rangle & |T_{-1,H}\beta_L\rangle & |S_H\beta_L\rangle \\
\hline
\langle T_{0,H}\beta_L| & -\omega_L & \frac{\pi J_{HL}}{\sqrt{2}} & \frac{\pi J_{HL}}{2} \\
\langle T_{-1,H}\beta_L| & \frac{\pi J_{HL}}{\sqrt{2}} & -\frac{2\pi J_{HL}}{4} - \omega_H & -\frac{\pi J_{HL}}{\sqrt{2}} \\
\langle S_H\beta_L| & \frac{\pi J_{HL}}{2} & -\frac{\pi J_{HL}}{\sqrt{2}} & -2\pi J_{HH} - \omega_L
\end{array}$$

$$\langle T_{-1,H}\beta_L| \quad \left(\frac{2\pi J_{HL}}{4} - \omega_H - \omega_L\right)$$

with basis $|T_{-1,H}\alpha_L\rangle$.

Alternatively, the AA'BB' system breaks out into two 1x1, 5 2x2, and one 4x4 subspaces in a basis set with singlet triplet on the hydrides and singlet triplet on the target nuclei shown here:

$$|T_{1,H}T_{1,L}\rangle$$
$$\langle T_{1,H}T_{1,L}| \quad (\pi J_{HL} + \omega_H + \omega_L)$$

$$\begin{array}{cc} & |T_{1,H}T_{0,L}\rangle \quad |T_{0,H}T_{1,L}\rangle \\ \langle T_{1,H}T_{0,L}| & \begin{pmatrix} \omega_H & \pi J_{HL} \\ \pi J_{HL} & \omega_L \end{pmatrix} \\ \langle T_{0,H}T_{1,L}| & \end{array}$$

$$\begin{array}{c} \phantom{xx} \\ \langle T_{1,H}T_{-1,L}| \\ \langle T_{0,H}T_{0,L}| \\ \langle T_{-1,H}T_{1,L}| \\ \langle S_H S_L| \end{array} \begin{pmatrix} |T_{1,H}T_{-1,L}\rangle & |T_{0,H}T_{0,L}\rangle & |T_{-1,H}T_{1,L}\rangle & |S_H S_L\rangle \\ -\pi J_{HL} + \omega_H - \omega_L & \pi J_{HL} & 0 & -\pi J_{HL} \\ \pi J_{HL} & 0 & \pi J_{HL} & \pi J_{HL} \\ 0 & \pi J_{HL} & -\pi J_{HL} - \omega_H + \omega_L & -\pi J_{HL} \\ -\pi J_{HL} & \pi J_{HL} & -\pi J_{HL} & -2\pi J_{HH} - 2\pi J_{LL} \end{pmatrix}$$

$$\begin{array}{cc} & |T_{1,H}S_L\rangle \quad |S_H T_{1,L}\rangle \\ \langle T_{1,H}S_L| & \begin{pmatrix} -2\pi J_{LL} + \omega_H & \pi J_{HL} \\ \pi J_{HL} & -2\pi J_{HH} + \omega_L \end{pmatrix} \\ \langle S_H T_{1,L}| & \end{array}$$

$$\begin{array}{cc} & |T_{0,H}T_{-1,L}\rangle \quad |T_{-1,H}T_{0,L}\rangle \\ \langle T_{0,H}T_{-1,L}| & \begin{pmatrix} -\omega_L & \pi J_{HL} \\ \pi J_{HL} & -\omega_H \end{pmatrix} \\ \langle T_{-1,H}T_{0,L}| & \end{array}$$

$$\begin{array}{cc} & |T_{0,H}S_L\rangle \quad |S_H T_{0,L}\rangle \\ \langle T_{0,H}S_L| & \begin{pmatrix} -2\pi J_{LL} & \pi J_{HL} \\ \pi J_{HL} & -2\pi J_{HH} \end{pmatrix} \\ \langle S_H T_{0,L}| & \end{array}$$

$$\begin{array}{cc} & |T_{-1,H}S_L\rangle \quad |S_H T_{-1,L}\rangle \\ \langle T_{-1,H}S_L| & \begin{pmatrix} -2\pi J_{LL} - \omega_H & \pi J_{HL} \\ \pi J_{HL} & -2\pi J_{HH} - \omega_H \end{pmatrix} \\ \langle S_H T_{-1,L}| & \end{array}$$

$$|T_{-1,H}T_{-1,L}\rangle$$
$$\langle T_{-1,H}T_{-1,L}| \quad (\pi J_{HL} - \omega_H - \omega_L)$$

Optimal Field Dependence on Exchange Rates

Figure 4 in the main manuscript uses a truncated 3 spin system to demonstrate that even the simplest spin systems are poorly approximated by a two-level system. This discrepancy is also seen in a 4 spin system shown in Figure S1

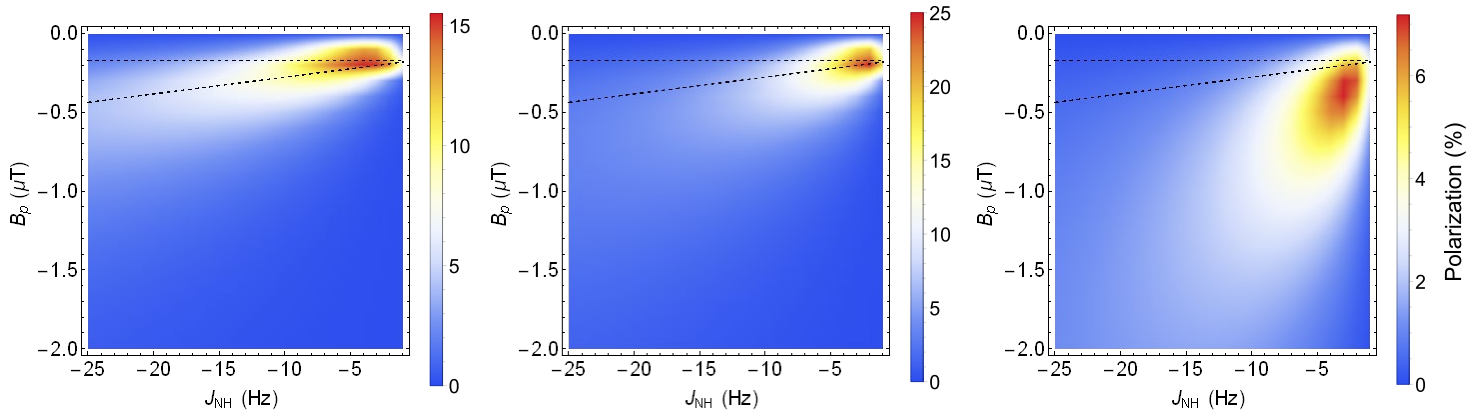

**Fig S1. 4 spin discrepancies between optimal polarization transfer fields and LAC matching conditions in continuous field experiments.** (A) Polarization after application of 10 seconds at a given field $B_p$ under a variety of $J_{NH}$ couplings and $k_L = 1\ s^{-1}$, (B) $k_L = 10\ s^{-1}$, and (C) $k_L = 100\ s^{-1}$. Black dashed lines correspond to the AA'BB' LAC conditions $\Delta\omega_{HL} = -2\pi(J_{HH} + J_{LL}) + \pi J_{HL}$ and $\Delta\omega_{HL} = -2\pi(J_{HH} - J_{LL})$. Additional simulation parameters: $[catalyst]:[ligand] = 1:20$, $k_H = 1\ s^{-1}$.

The 3 spin and 4 spin systems presented so far are simplified approximations of experimental systems. Running a full acetonitrile spin system including coupling out to the methyl protons shows results consistent with the standard 3 spin system.

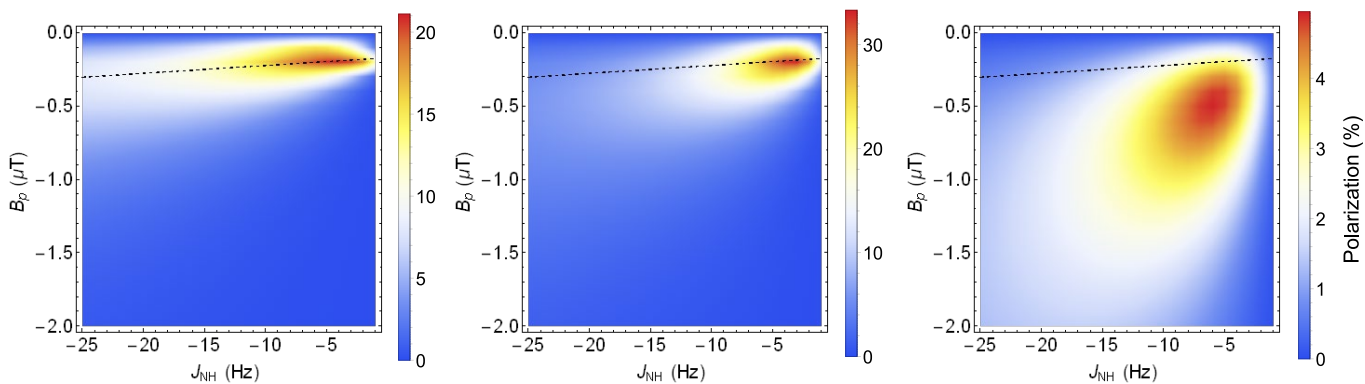

**Fig S2. Full acetonitrile spin system simulations showing discrepancies between optimal polarization transfer fields and LAC matching conditions in continuous field experiments.** (A) Polarization after application of 10 seconds at a given field $B_p$ under a variety of $J_{NH}$ couplings and $k_L = 1\ s^{-1}$, (B) $k_L = 10\ s^{-1}$, and (C) $k_L = 100\ s^{-1}$. Black dashed lines correspond to the AA'B LAC condition $\Delta\omega_{HL} = 2\pi J_{HH} - \pi J_{NH}/2$. Additional simulation parameters: $[catalyst]:[ligand] = 1:20$, $k_H = 1\ s^{-1}$.

We also see the same deviation from predicted optima under the LAC matching condition for coherently pumped SABRE SHEATH when simulating a full experimental acetonitrile system. Because the numerical simulations assume perfect homogeneity, inhomogeneity during the delay period must be introduced to ensure that slowly exchanging systems ($k_L = 1 s^{-1}$) we don't see phase shifting effects from action of the $\Delta\omega_{HL}$ term during the 350 ms long delay period ($\theta \gg 2\pi$). These effects would require exceptional field homogeneity and would be unlikely to impact real experimental systems, so artificial inhomogeneities of ±5% were added to the simulations shown in Figure S3.

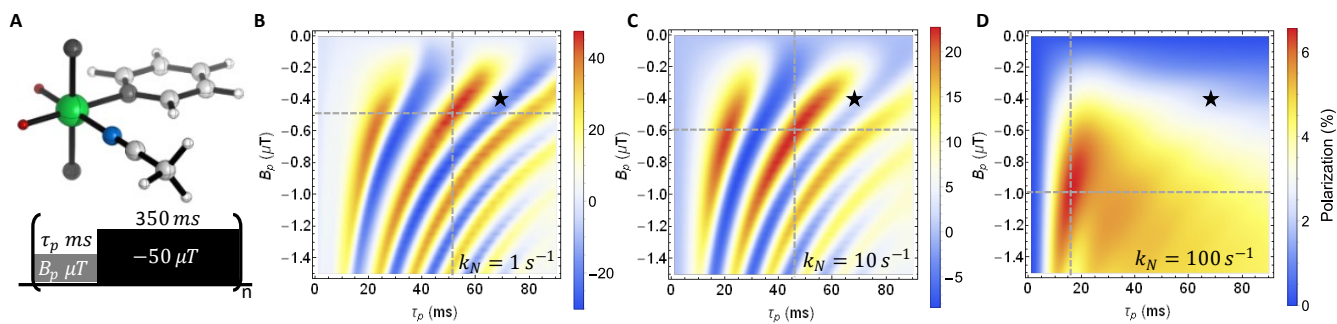

**Fig S3. Discrepancies between optimal polarization transfer fields and LAC matching conditions in coherently pumped SABRE SHEATH experiments on a full $^{15}$N-acetonitrile spin system.** (A) Full acetonitrile spin system and the pulse sequence used in the simulations. (B) Polarization after application of the pulse sequence in A for 30 seconds under a variety of evolution field conditions $B_p$ and durations $\tau_p$ and $k_N = 1\ s^{-1}$, (C) $k_N = 10\ s^{-1}$, and (D) $k_N = 100\ s^{-1}$. Dashed gray lines indicate maximum polarization conditions. Stars indicate optima predicted using LAC matching condition. Additional simulation parameters: $[catalyst]:[ligand] = 1:5$, $k_H = 1\ s^{-1}$, inhomogeneity: 5%

Coherence Transfer Pathways

The coherence transfer pathway leading from singlet order on the hydrides to magnetization on the target nucleus goes through 4 commutators starting with initial density matrix:

$$\hat{\rho}_0 = \left(\frac{1}{4}\hat{1} - \hat{I}_1 \cdot \hat{I}_2\right) \otimes \hat{1}$$

And rearranged nuclear spin Hamiltonian:

$$\hat{\mathcal{H}}' = 2\pi J_{HH}\hat{I}_1 \cdot \hat{I}_2 + 2\pi J_{HL}\hat{I}_{1z}\hat{L}_z - \Delta\omega_{HL}\hat{L}_z + 2\pi J_{HL}(\hat{I}_{1x}\hat{L}_x + \hat{I}_{1y}\hat{L}_y)$$

Starting with the first order term in the series, $it[\hat{\rho}_0, \hat{\mathcal{H}}]$, we rotate the initial density matrix under the $J_{NH}$ term as $\hat{\rho}_o$ commutes with all other terms in the Hamiltonian:

$$\dot{\rho}t = it[\hat{\rho}_o, \hat{H}] = it[\hat{I}_1 \cdot \hat{I}_2, 2\pi J_{HL}(\hat{I}_{1x}\hat{L}_x + \hat{I}_{1y}\hat{L}_y + \hat{I}_{1z}\hat{L}_z)]$$
$$= 2\pi J_{LH}t(-\hat{I}_{1z}\hat{I}_{2x}\hat{L}_y + \hat{I}_{1y}\hat{I}_{2x}\hat{L}_z + \hat{I}_{1z}\hat{I}_{2y}\hat{L}_x - \hat{I}_{1x}\hat{I}_{2y}\hat{L}_z - \hat{I}_{1y}\hat{I}_{2z}\hat{L}_x + \hat{I}_{1x}\hat{I}_{2z}\hat{L}_y)$$

The important second and third order terms arise from rotations about the $J_{HH}$ and $\Delta\omega_{HL}$ terms. Because these terms commute with one another, the order of commutation is unimportant. The second order terms are then:

$$\ddot{\rho}\frac{t^2}{2!} = i\frac{t^2}{2!}[\dot{\rho}, -\Delta\omega_{HL}\hat{L}_z] + i\frac{t^2}{2!}[\dot{\rho}, 2\pi J_{HH}\hat{I}_1 \cdot \hat{I}_2]$$
$$= 4\pi J_{HL}(-\Delta\omega_{HL})\frac{t^2}{2!}(\hat{I}_{1z}\hat{I}_{2y}\hat{L}_y + \hat{I}_{1z}\hat{I}_{2x}\hat{L}_x - \hat{I}_{1x}\hat{I}_{2z}\hat{L}_x - \hat{I}_{1y}\hat{I}_{2z}\hat{L}_y)$$
$$+ 2\pi^2 J_{HL}J_{HH}\frac{t^2}{2!}(\hat{I}_2 \cdot \hat{L} - \hat{I}_1 \cdot \hat{L})$$

and third order terms are:

$$\dddot{\rho}\frac{t^3}{3!} = i\frac{t^3}{3!}[\ddot{\rho}, -\Delta\omega_{HL}\hat{L}_z] + i\frac{t^3}{3!}[\ddot{\rho}, 2\pi J_{HH}\hat{I}_1 \cdot \hat{I}_2] = 8\pi^2 J_{HL}(-\Delta\omega_{HL})J_{HH}\frac{t^3}{3!}(\hat{I}_{1x}\hat{L}_y - \hat{I}_{2x}\hat{L}_y - \hat{I}_{1y}\hat{L}_x + \hat{I}_{2y}\hat{L}_x)$$

The final commutator goes through the transverse terms in the $J_{HL}$ coupling to result in direct magnetization on the target nucleus $\hat{L}_z$ in the fourth order term with a $t^4$ dependence:

$$\ddddot{\rho}\frac{t^4}{4!} = i\frac{t^3}{3!}[\dddot{\rho}, 2\pi J_{HL}(\hat{I}_{1x}\hat{L}_x + \hat{I}_{1y}\hat{L}_y)] = 8\pi^3 J_{HL}^2 \Delta\omega_{HL}J_{HH}\frac{t^4}{4!}(2\hat{L}_z - 2\hat{I}_z)$$

Application of Field Sequences

The zero-field condition for experimental zero field pulses was found by taking a standard continuous field experiment magnetic field profile. The applied voltage was swept through from 0.2V to -0.2V across a 118Ω resistor in series with a solenoid ($N = 184, \ell = 21\ 1/8"$). This voltage was then used for the zero-field condition. The power source for all experiments was a BK Precision 4053 10 MHz Arbitrary Waveform Generator. Field sequences were applied and then prior to removing the sample from the $\mu$-metal shield, the power supply was switched to a constant 5V supplied by a GW Instek GPS-3030D Laboratory DC Power Supply to ensure that the solenoid field was parallel to the lab z field. This instantaneous field switch prevents exposure of the sample to an adiabatic field sweep through the resonance condition when removing the sample from the $\mu$-metal shield for measurement. Applied voltages were confirmed via oscilloscope trace with a Picoscope 2205A (25 MHz 8 bit 250 MS/s). Figure S3 shows one example trace for a compensated and uncompensated pulse sequence.

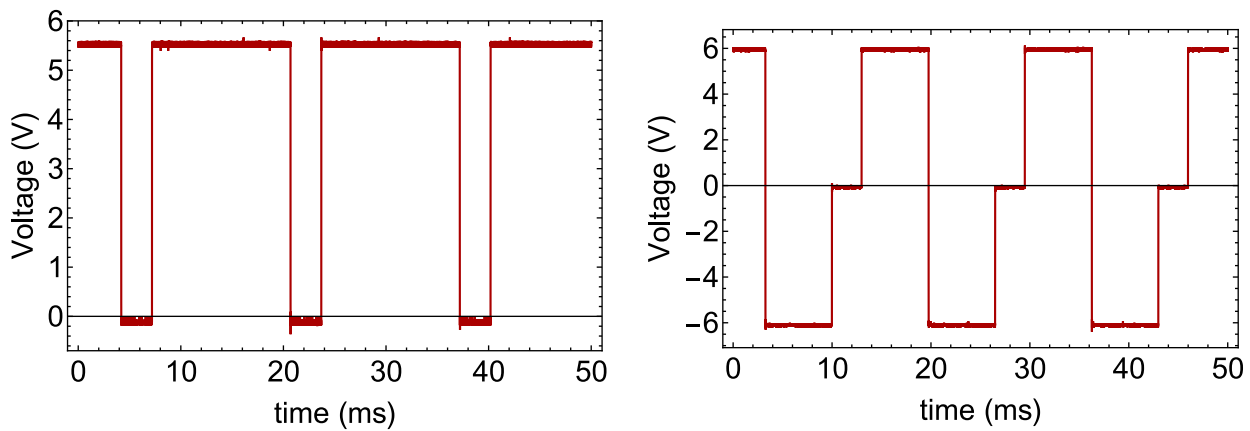

**Fig S4. Voltage Traces for inhomogeneity uncompensated and compensated zero-field pulse sequences.** (A) Oscilloscope voltage trace for a zero-field sequence ($\tau_p = 3\ ms$, $B_p = 0\ \mu T$, $B_d = -22.35\ \mu T$, $\tau_d = 13.5\ ms$). (B) Oscilloscope trace for inhomogeneity compensated zero-field sequence ($\tau_p = 3\ ms$, $B_p = 0\ \mu T$, $B_d = \pm 22.35\ \mu T$, $\tau_d = 13.5\ ms$, $offset = 0\ ms$).

## Zero field pulse duration sweep

The duration of the zero-field pulse was demonstrated to be important for determining the net phase rotation imparted on $J_{HL}$ over the course of the lifetime of a complex. This parameter will be optimized to different extents based on the homogeneity and accuracy of the zero-field pulse, the duration and accuracy of the high field pulse, and the exchange rate of the PTC of interest. After optimizing all other parameters in the system with much narrower peaks, we also optimized the zero-field pulse duration $\tau_p$ and the results are shown in Figure S5. A peak enhancement was found at 1.5 ms, but the optima is broad relative to the higher field pulse magnitude and duration.

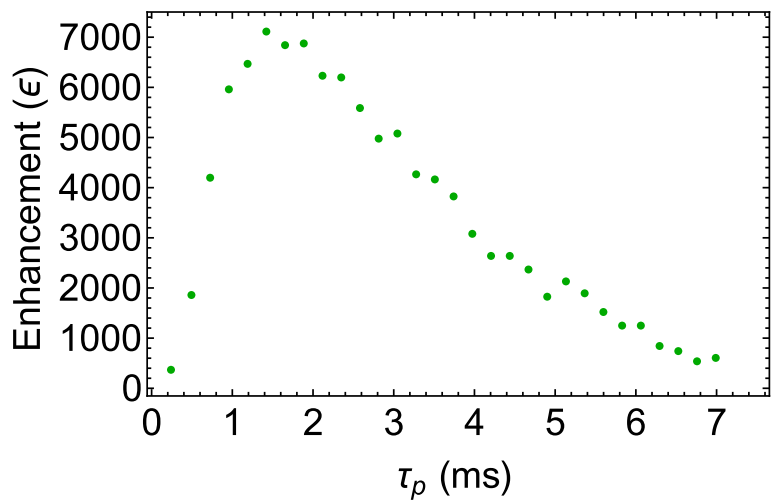

**Fig S5. Zero-field duration sweep.** Experimental enhancement of $^{15}$N over thermal polarization in a 1T NMR taken on an experimental Ir(H)$_2$(IMes)(pyr)$_2$($^{15}$N-acetonitrile) spin system exposed to a zero-field sequence with $B_p = 0\ \mu T$, $\tau_d = 13.5\ ms$, $B_d = -22.35\ \mu T$ (sample composition: 5 mM Ir(IMes), 50 mM $^{14}$N-pyridine, 100 mM $^{15}$N-acetonitrile in CD$_3$OD)

Making Field Measurements Using Oscillations

For labs without a precise magnetometer which functions in the $nT - \mu T$ regime, the inherent oscillations present in these experiments can be used to determine exact field measurements. In a $B_d$ field sweep for the high field magnitude in a zero-field/high-field pulse sequence, the zero crossings occur every $\pi n = 2\pi B_d \tau_d (\gamma_H - \gamma_N)$. This can also be written using some coefficient $A$ converting between the applied current $I_d$ and the field $B_d$: $\pi n = 2\pi (AI_d)\tau_d(\gamma_H - \gamma_N)$. Measurement of the applied current difference between zero-crossings gives a measure of $I_d$ and we may solve for the exact conversion factor:

$$A = \frac{1}{2I_d \tau_d (\gamma_H - \gamma_N)}$$

This provides a precise measure of the field generated by any given solenoid in the $\mu T$ regime.

Impact of Inhomogeneity

The features explored in this paper are sharp and require a homogenous field environment to resolve (aside from the inhomogeneity compensated pulse sequence in Fig. 10). For a zero-field/high-field experiment with a fixed (but inhomogenous) $B_d$, sweeping through the $\tau_d$ can clearly demonstrate the impact of inhomogeneity on the experiment. At longer times $\tau_d$, there have been many cycles through $\Delta\omega_{HL}$ and any inhomogeneities present in the field will result in larger deviations from the sharp optimal condition resulting in a reduction in polarization at longer times $\tau_d$ as shown in Figure S5. This is problematic because longer times and larger $\theta$ are needed to reduce the effective $J_{HL}$ coupling into a more optimal regime.

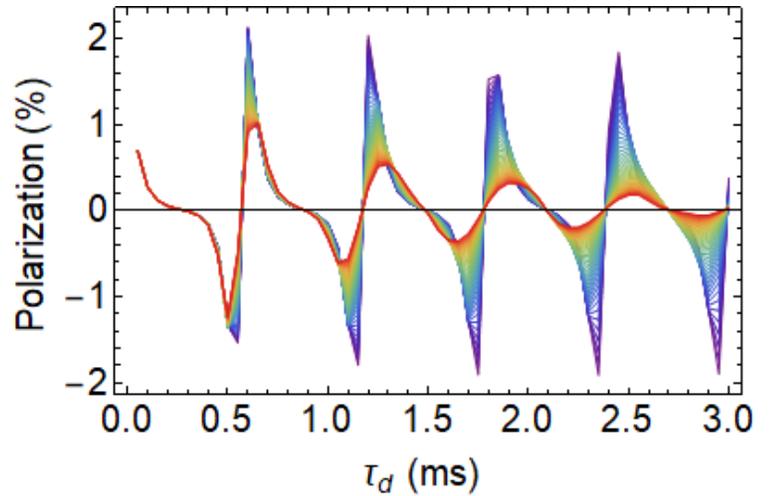

**Fig S6. Inhomogeneity causing reduction in 2 State SABRE enhancements.** 3 spin SABRE simulations with $B_p = -0.5\ \mu T$, $\tau_p = 3\ ms$, $B_d = -40\ \mu T$, and variable $\tau_d$ with different degrees of inhomogeneities present (red = 5%, purple = 0.5%). Wider distributions of magnetic fields results in peak flattening and suppresses polarization build-up under these conditions.

Polarization dependence on exchange rates

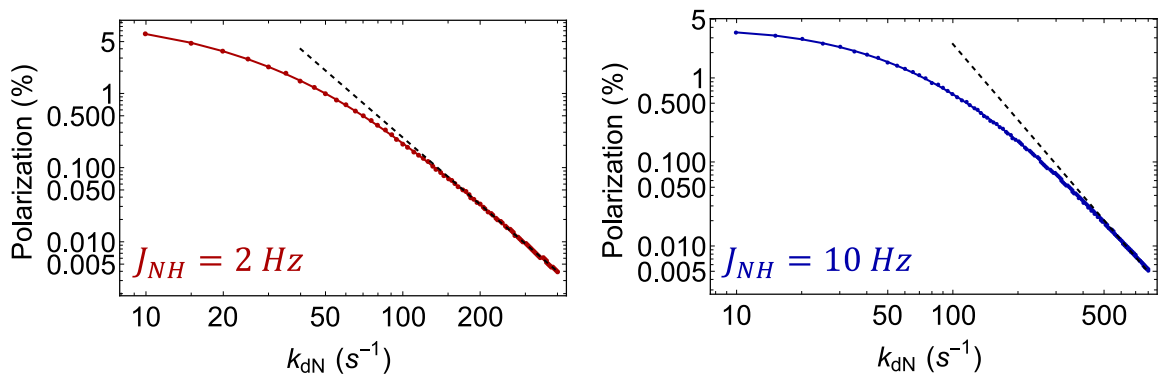

**Fig S7. Final polarization dependence on exchange rate.** Final polarization of 15N in a 3 spin SABRE system after 10 seconds of exposure to the LAC matching condition field (B = $(2\pi J_{HH} - (\pi J_{NH})/2)/(\gamma_H - \gamma_N)$). At high exchange rates we see a $k^{-3}$ dependence (dashed black line) for the polarization build-up as expected and discussed in the main text. $J_{NH} = 2\ Hz$ is shown on the left and $J_{NH} = 10\ Hz$ is shown on the right, both demonstrating this same $k^{-3}$ dependence at high exchange rates.